\documentclass[prb,twocolumn,showpacs]{revtex4-1}
	\usepackage{graphicx,epsf,bm,bbm,color,amsmath,amssymb}
\usepackage{ulem}

	\newcommand{\be}{\begin{equation}}
	\newcommand{\beqn}{\begin{eqnarray}}
	\newcommand{\ee}{\end{equation}}
	\newcommand{\eeqn}{\end{eqnarray}}

	\newcommand{\bsq}{{\boldsymbol q}}
	
	\renewcommand{\k}{{\bf k}}
	\renewcommand{\a}{{\bf a}}
	\newcommand{\q}{{\bf q}}
	\newcommand{\K}{{\bf K}}
	\newcommand{\M}{{\bf M}}
	\newcommand{\R}{{\bf R}}

\newcommand{\dd}{{\boldsymbol \delta}}

\begin{document}

		\title{An equivalence between monolayer and bilayer honeycomb lattices}
		\author{ G. Montambaux}
		\affiliation{Laboratoire de Physique des Solides, CNRS UMR 8502, Univ. Paris-Sud, F-91405 Orsay cedex, France.\\}
		\begin{abstract}
In this brief report, we show the equivalence between the tight-binding descriptions of the monolayer and bilayer honeycomb lattices. With appropriate value of the third nearest neighbors coupling, the Hamiltonian for a monolayer is equivalent to the low energy effective Hamiltonian for bilayer in the presence of trigonal warping. A simple physical argument is provided to explain this correspondance.

		\end{abstract}
	\maketitle
	
	\section{Introduction}

Although the literature on physical properties of monolayer and bilayer graphene  based on the tight-binding description of the electronic spectrum is quite large,\cite{Rev} it may be useful to precise a few simple relations between these two systems.
In this brief report,  we consider the tight-binding Hamiltonian on the monolayer honeycomb lattice with coupling  between nearest neighbors ($t$) and third nearest neighbors $(t_3)$. It was noticed that for a critical value of the coupling $t_3$ ($=t/2)$, the low energy spectrum is quadratic, similar to the case of bilayer graphene.\cite{Bena2} It turns out that the low energy spectrum in bilayers is not exactly quadratic, due to a small additional hopping contribution between the layers leading to   a trigonal warping of the spectrum.\cite{MF06}  Here we show that close to the  critical value of the coupling $t_3$  $( \lesssim t/2)$, there is an exact correspondance with  the low energy Hamiltonian for bilayer graphene in the presence of trigonal warping.
We show that  there is also a simple correspondance between anisotropic monolayer graphene and bilayer graphene in the presence of a translation between the layers.
This report is organized as follows:
 In the next section, we introduce the tight-binding Hamiltonian for a monolayer and consider its limits when $t_3 \gtrsim t/3$ and  $t_3 \lesssim t/2$ corresponding respectively to the emergence of new Dirac points in the vicinity of the three inequivalent $\M$ points of the reciprocal space and to their merging at the $\K$ points. In section \ref{sect.bilayer}, we recall the effective low energy  Hamiltonian for bilayer graphene and we discuss its relation with the monolayer Hamiltonian, with a correspondance between the relevant parameters. We briefly conclude in the last section and we stress that  a simple model with a single parameter $t_3/t$ allows for  the continuous description of  the evolution between different scenarios of merging (or emergence) of Dirac points with opposite or same Berry phase.  These remarks provide a simple tool to describe in a coherent picture a large range of different physical situations.

\section{Monolayer with third neighbor coupling}

\begin{figure}[h!]
\begin{center}
\includegraphics[width=6cm]{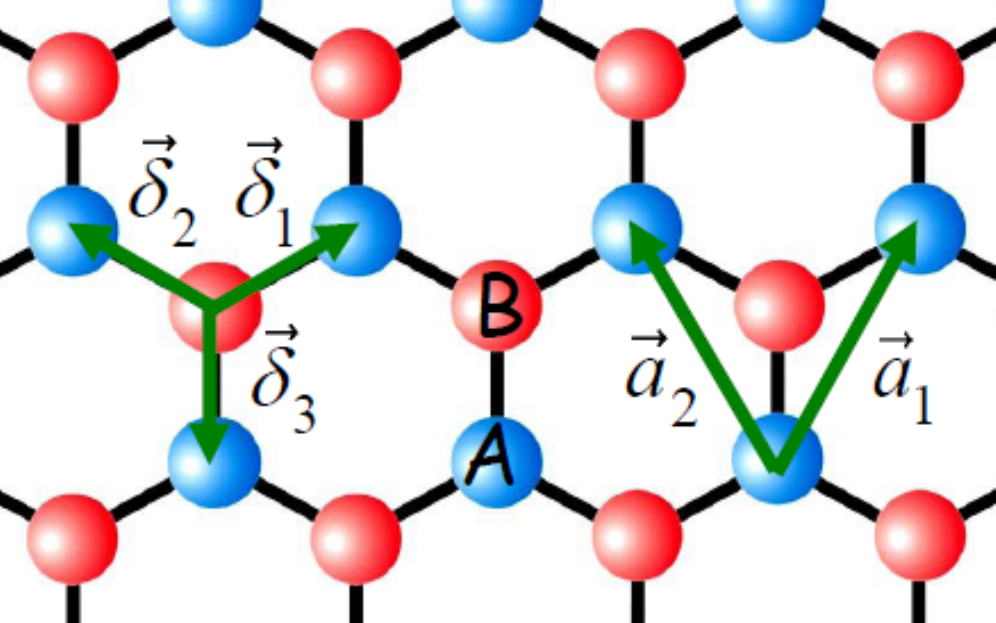}
\end{center}
\caption{Honeycomb lattice. The $\dd_i$ connect one atom to its nearest neighbors. The $\a_i$ are Bravais lattice vectors.}
\label{fig:monolayer}
\end{figure}

Let us first recall the tight-binding Hamiltonian for the monolayer honeycomb lattice~:
\be {\cal H}_{mono} =  - t  \sum_{\langle i,j \rangle} (a^\dagger_i b_j + h.c. )-  t_3  \sum_{\langle\langle\langle i,j \rangle\rangle\rangle} (a^\dagger_i b_j + h.c. )  \ . \label{Hmono} \ee
where $a_i$ ($a^\dagger_i$) annihilates (creates) and electron on a $A$ site and there is a similar definition of the $b_i$ operators on the $B$ sites (Fig. \ref{fig:monolayer}). The first term  describes the coupling between nearest neighbors    ($t$ being the hopping amplitude), and the second term describes the coupling between third nearest neighbors ($t_3$ being the hopping amplitude). These two terms correspond to coupling between sites of different sublattices. We have not considered here the coupling between second nearest neighbors (which couples sites of the same sublattice), because it is not relevant here for our purpose.
This Hamiltonian can be rewritten in the form  ${\mathcal H}= \sum_\k \psi^\dagger_\k  \, {\mathcal H}(\k) \, \psi_\k$ with $\psi^\dagger_\k =(a^\dagger_\k, b^\dagger_\k)$ and\cite{remark1}

\be {\cal H}_{mono}(\k) =  -\left(
                  \begin{array}{cc}
                    0 &  t f(\k) + t_3 f_3(\k)   \\
                  t f^*(\k) + t_3 f_3^*(\k) & 0 \\
                  \end{array}
                \right)
                \label{Hmono2}
               \ee
where the functions $f(\k)$ and $f_3(\k)$ are given by

                \be f(\k)=e^{i \k . \dd_1} + e^{i \k . \dd_2}+e^{i \k . \dd_3} \ee
                 \be f_3(\k) =e^{-2 i \k . \dd_1} + e^{-2 i \k . \dd_2}+e^{-2 i \k . \dd_3} \,  \ee
and the vectors $\dd_i$ connect one atom to its three nearest neighbors (Fig. \ref{fig:monolayer}):
\be \dd_1=a({\sqrt{3} \over 2},{1 \over 2}) \  , \quad \dd_2=a(-{\sqrt{3} \over 2},{1 \over 2}) \  , \quad \dd_3=a(0,-1)  \ , \ee
$a$ being the interatomic distance.

\begin{figure}[h!]
\begin{center}
\includegraphics[width=4.5cm]{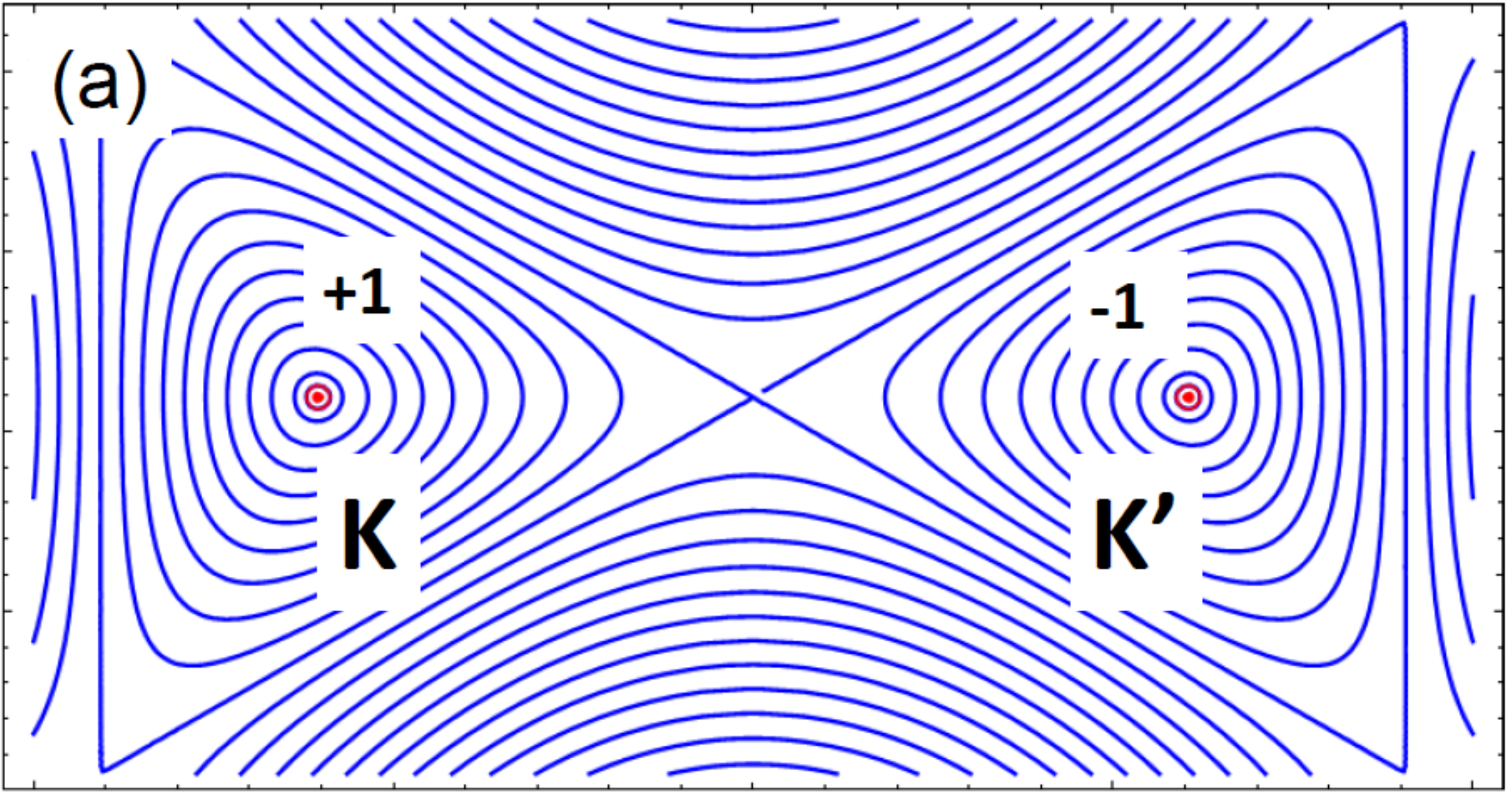}\includegraphics[width=4.5cm]{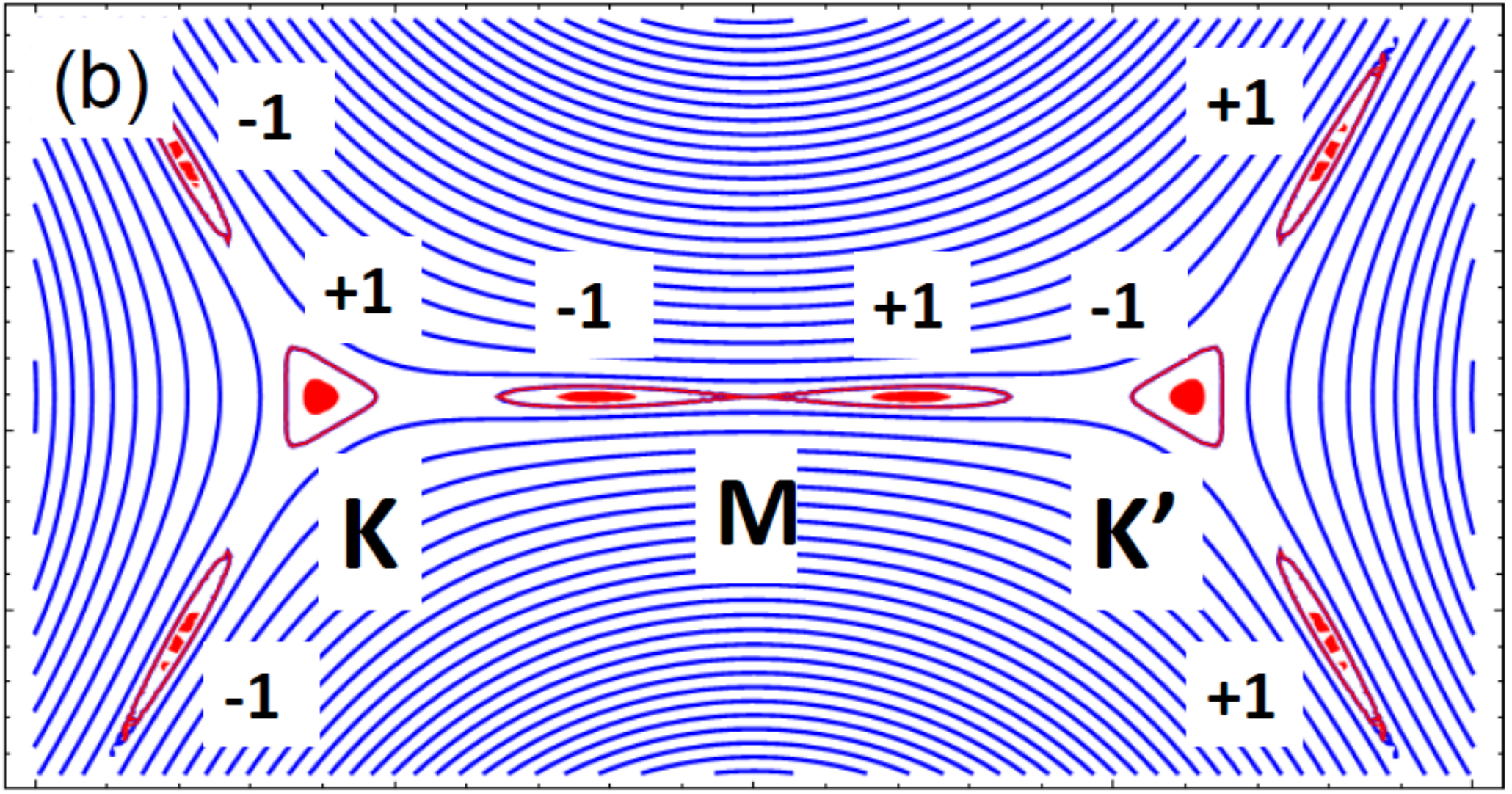}
\end{center}
\begin{center}
\includegraphics[width=4.5cm]{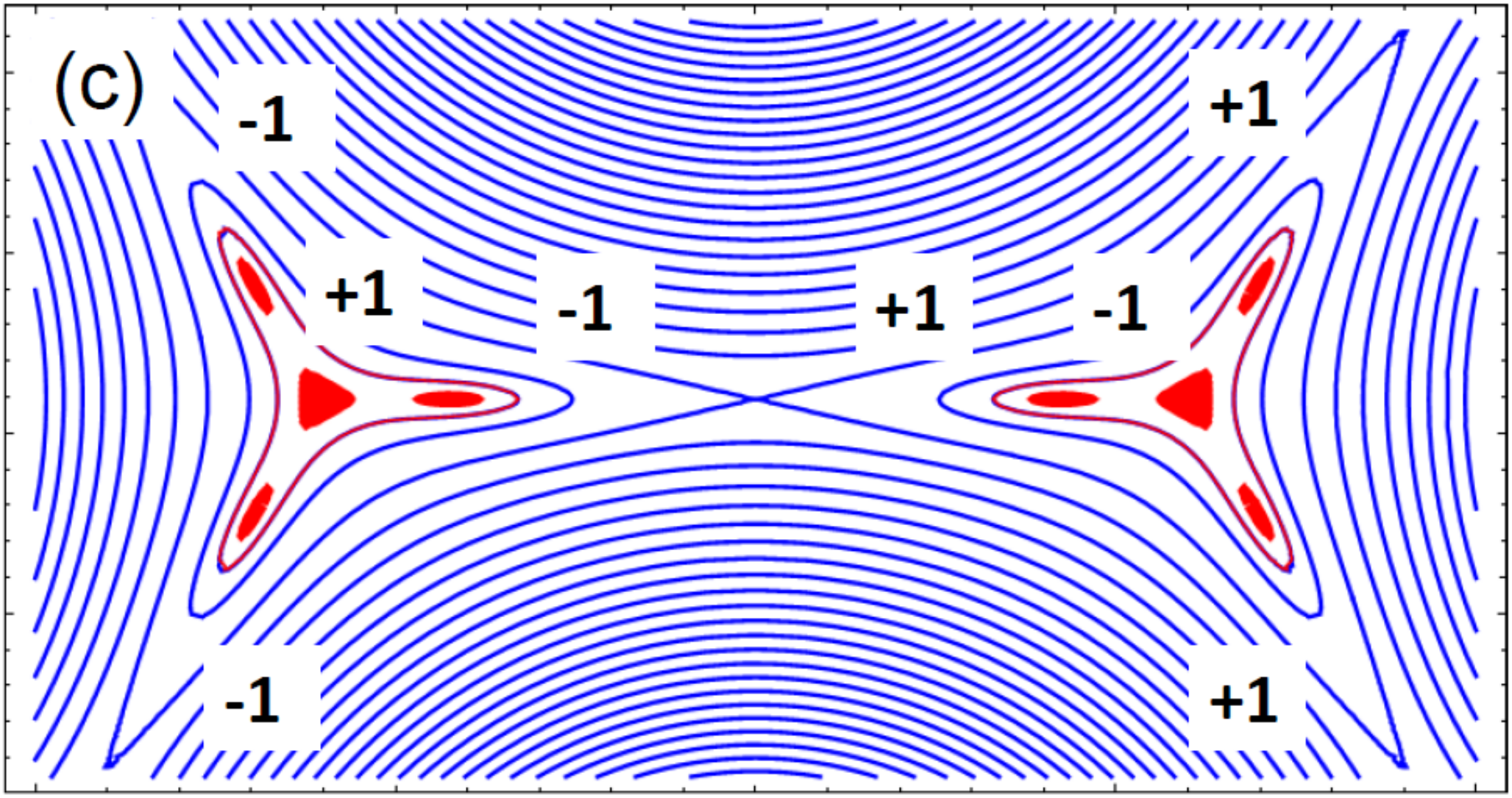}\includegraphics[width=4.5cm]{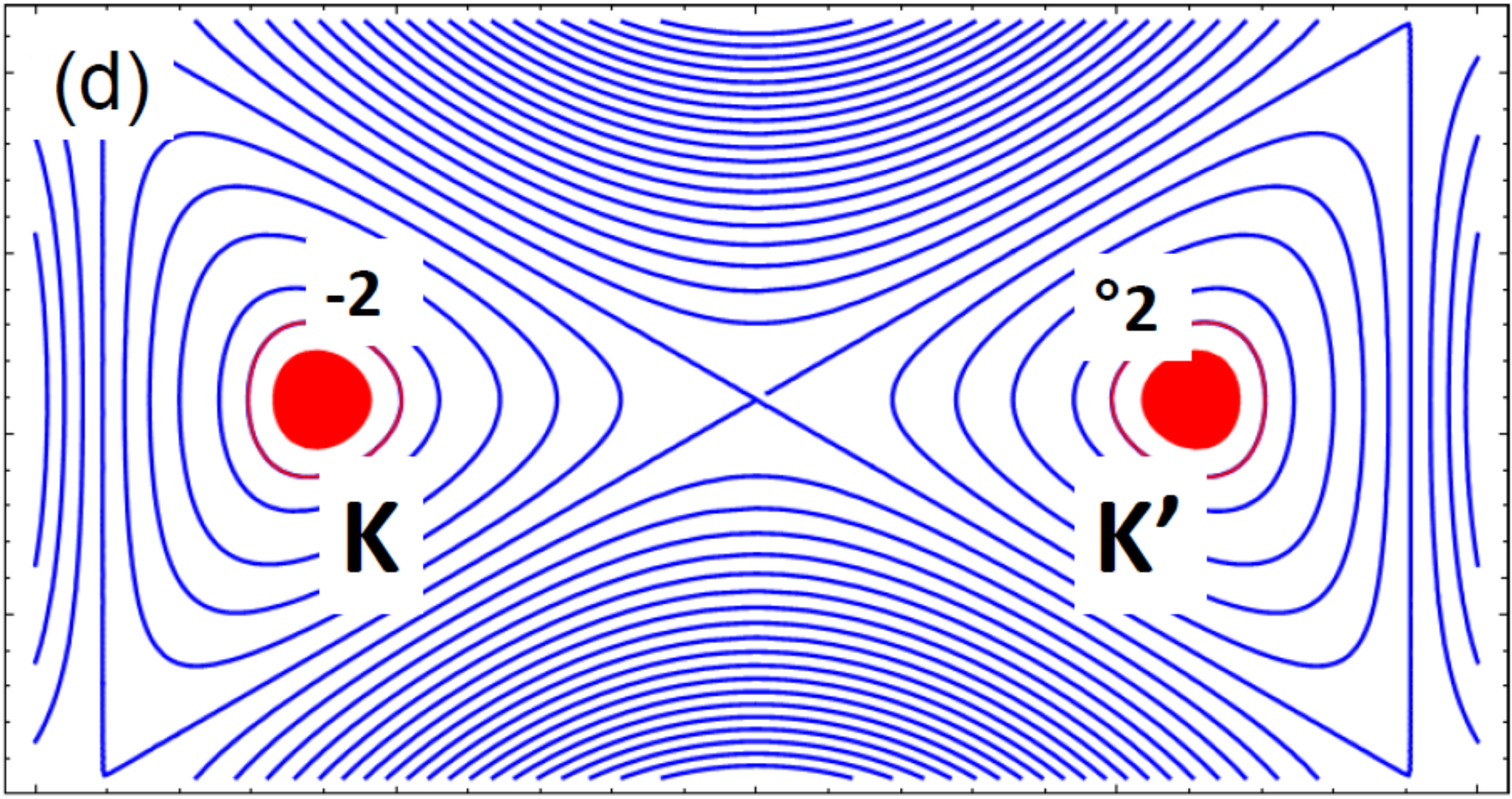}
\end{center}
\begin{center}
\includegraphics[width=4.5cm]{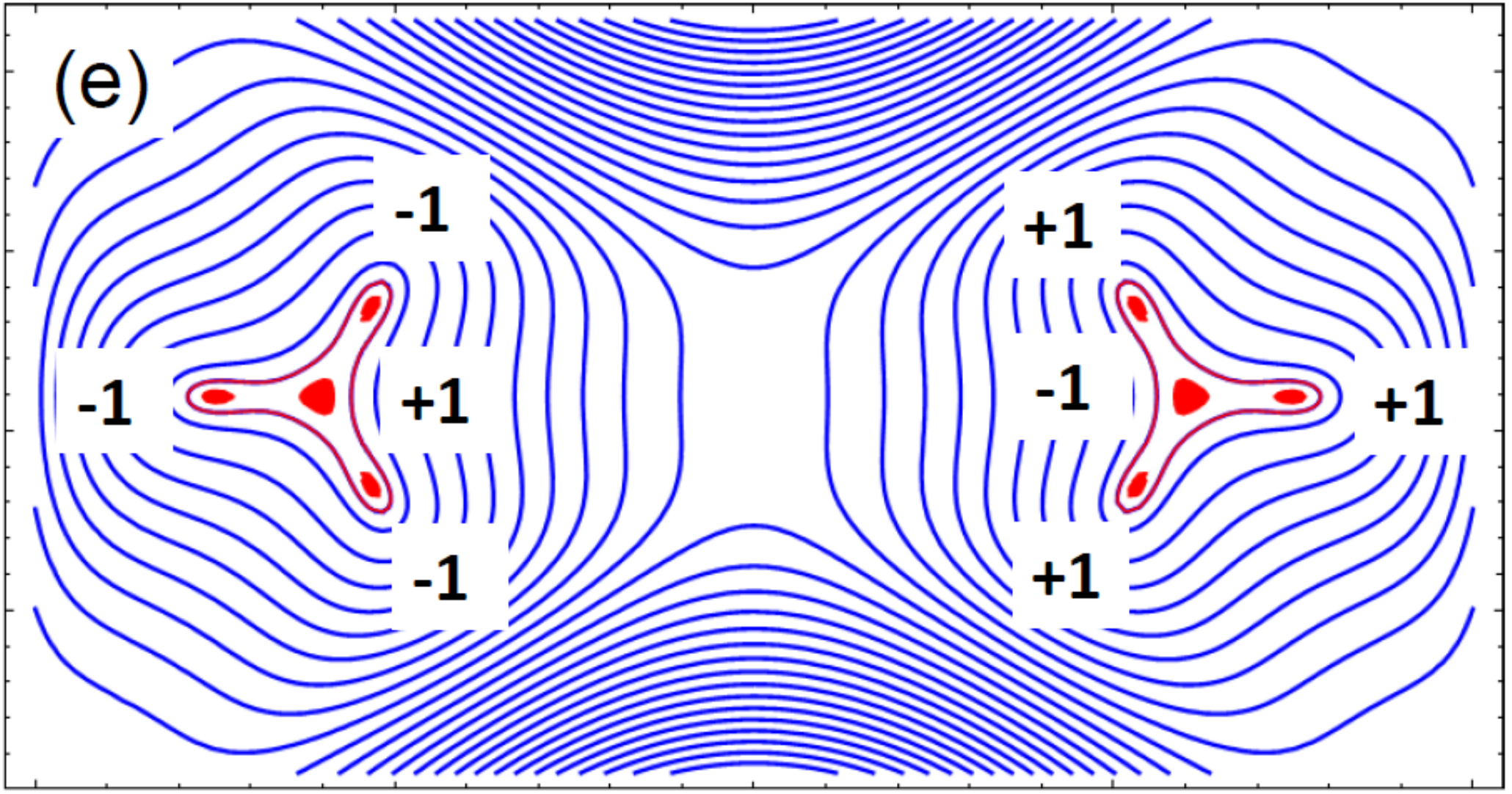}
\end{center}
\caption{Iso-energy lines in the vicinity of the $\K^{(')}$ and $\M$ points, for different values of the parameter $t_3$. (a) $t_3 = 0$, (b) $t_3 = 0.35 t$, (c) $t_3 = 0.40 t$, (d) $t_3 = 0.5 t$, , (e) $t_3 = 0.65 t$. The vicinity of the Dirac points is indicated in red, as well as their associated Berry phase (within a factor $\pi$).}
\label{fig:t3}
\end{figure}

In graphene, the $t_3$ term is small.\cite{reich} However, it is of interest to imagine a larger value of this parameter because it has a quite interesting effect on the evolution of the spectrum, as has been theoretically considered in ref.[\onlinecite{Bena2}].
 When $t_3$ increases and reaches the critical value $t/3$, a new pair of Dirac points emerges from each of the three inequivalent $\M$-point of the reciprocal space (see Fig. \ref{fig:t3}.a,b), following the universal scenario predicted for the apparition of Dirac points in  a $2D$ crystal in the vicinity of time-reversal symmetry points.\cite{Montambaux1,Montambaux2} Writing $\k = \k_M + \q$ , we obtain the following "universal Hamiltonian" in the vicinity of $t_3 =t/3$ (keeping the leading order terms)
 \be {\mathcal H}_{univ.}(\q)= \left(
                \begin{array}{cc}
                 0 &  \displaystyle  i c q_y + {q_x^2 \over 2 m^*} + \Delta_* \\
                 \displaystyle   -i c q_y + {q_x^2 \over 2 m^*} + \Delta_* & 0 \\
                \end{array}
              \right) \label{HU} \ee
  where $\q=(q_x,q_y)$ and             where the parameters $m^*, c, \Delta_*$ can be related to the original band parameters. Here we find  $\Delta_*= t - 3 t_3 $,  $c= 2 t$ and  $m^* =2 / t$. The parameter $\Delta_*$, when becoming negative ($t_3 > t/3$), drives the emergence of a new pair of Dirac points at the $\M$-point (Fig. \ref{fig:t3}.b). The distance between the new Dirac points is given by $\Delta q = 2\sqrt{- 2 m^* \Delta^*}= 4 \sqrt{3 t_3/t -1}$

 When increasing further $t_3$, the new Dirac points approach the $\K$ and $\K'$ points, so that each initial Dirac point sitting at the $\K^{(')}$ points is surrounded by three Dirac points (with opposite Berry phases, see Fig. \ref{fig:t3}.c.).\cite{remark2} These Dirac points merge at the critical value $t_3 =t/2$, and the spectrum becomes quadratic around $\K^{(')}$, as already noticed in Ref. \cite{Bena2}. We will show in the next section that there is an exact correspondance between the Hamiltonian (\ref{Hmono}) with $t_3=t/2$ and the low energy Hamiltonian for bilayer graphene. When $t_3 \simeq t/2$, ${\mathcal H}_{mono}$ takes a new universal form (keeping leading order terms) in the vicinity of the $\K'$ point

 \be \mathcal{H}'_{univ}(\q)= \left(
                \begin{array}{cc}
                 0 & \displaystyle  -{\bsq^2 \over 2 m^*}+ c \, \bsq^\dagger + \Delta \\
                \displaystyle      -{{\bsq^\dagger}^2 \over 2 m^*}+ c \, \bsq + \Delta & 0 \\
                \end{array}
              \right) \label{Hunivprime} \ee
where $\bsq=q_x + i q_y$ (The Hamiltonian in  the vicinity of the $\K$ point is obtained by the substitution  $\bsq \rightarrow - \bsq^\dagger$). As we shall see in the next section, this is precisely the low energy effective Hamiltonian for sliding bilayer graphene.\cite{coreans,Falko11,Ecrys,Degail12}  Starting from (\ref{Hmono2}), we find $m^*= 4  / 9 t$ and $c=3  (t_3 - t/2) $ . Here the parameter $\Delta=0$. We comment on its effect in the next section. We see that the parameter $c$  controls  the number of Dirac points. When $c=0$, the spectrum is quadratic. When $c$ becomes finite, three Dirac points emerge in a trigonal arrangement.\cite{MF06} The orientation of this arrangement depends on the sign of $ m^* c$, that is the sign of $t (2 t_3 - t)$.

  \begin{figure}[h!]
\begin{center}
\includegraphics[width=7cm]{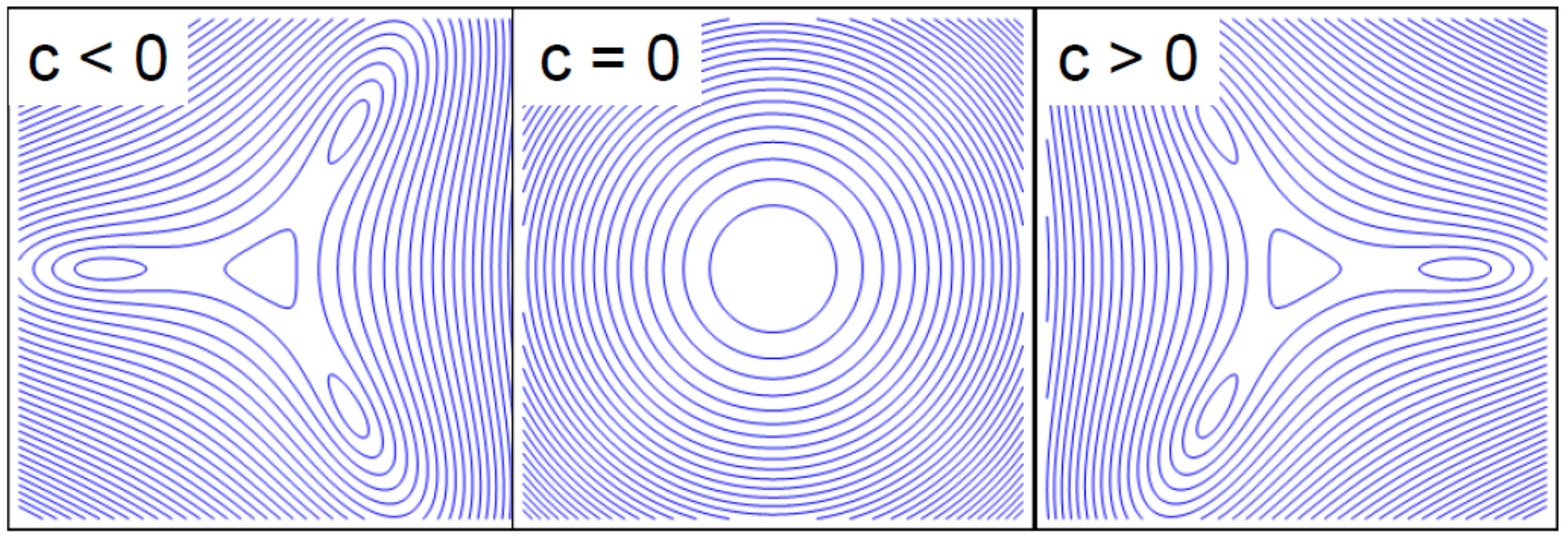}
\end{center}
\caption{Iso-energy lines for the Hamiltonian (\ref{Hunivprime}), here with $\Delta=0$. When $c=0$ the spectrum is quadratic. When $c \neq 0$, the spectrum splits into four Dirac points, keeping conserved the total Berry phase. The orientation of the Dirac points depends on the sign of $m^* c$. In changes when $t_3$ becomes larger than $t/2$. These figures (here for $m^* >0$) have to be compared with the spectrum in the vicinity of the $\K'$ point shown in Figs. \ref{fig:t3}.c-e for $t_3 \lesssim t/2$, $t_3 = t/2$ and $t_3 \gtrsim t/2$.}
\label{fig:variation-c}
\end{figure}

\section{Bilayer}
\label{sect.bilayer}

\begin{figure}[h!]
\begin{center}
\includegraphics[width=5cm]{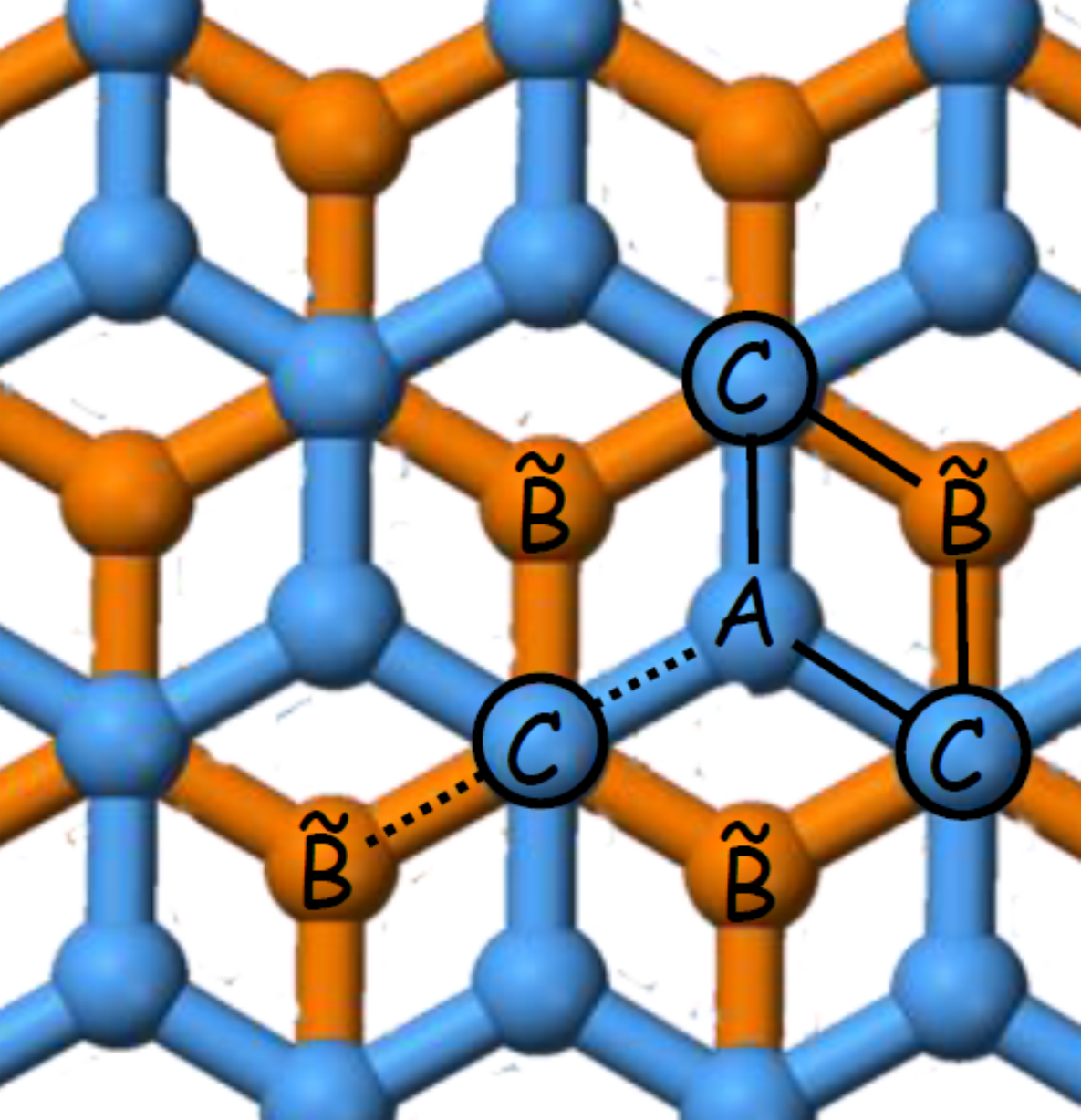}
\caption{Graphene bilayer in the Bernal stacking. The atoms ${\tilde A}$ and $B$ are on top of  each other. The pair (${\tilde A}$-$B$) is denoted $C$. This picture shows that the indirect coupling between $A$ and ${\tilde B}$ sites results from  second order processes where the electron jumps from $A$ to $C =({\tilde A}, B)$ and then from $C$ to ${\tilde B}$. There are two different processes indicated on the figure, one with black full lines, one with a dotted line. The first  process implies to possible paths and its amplitude is  thus twice larger than the second one.}
\label{fig:reseau-1}
\end{center}
\end{figure}

We now consider the Hamiltonian describing a bilayer of honeycomb lattices arranged in the so-called Bernal (or $A-B$) stacking (Fig. \ref{fig:reseau-1}. This Hamiltonian reads \cite{Rev}

\begin{eqnarray}  {\cal H}_{bi}  =&-& \gamma_0  \sum_{\langle i,j \rangle} (a^\dagger_i b_j + h.c. )  \nonumber \\
 &-& \gamma_0  \sum_{\langle i,j \rangle} ({\tilde a}^\dagger_i {\tilde b}_j + h.c. )  \nonumber \\
  &-& \gamma_1  \sum_{j} ({\tilde a}^\dagger_i  b_j + h.c. )  \nonumber \\
    &-& \gamma_3  \sum_{j} ({a}^\dagger_i  {\tilde b}_j + h.c. )   \ .  \\
    \end{eqnarray}
 The $a_i, b_i$ operate on one layer ($A$ and $B$ sites) and the ${\tilde a}_i, {\tilde b}_i$ operate on the other layer. The  $\gamma_0$ terms describe the coupling between nearest neighbors in each layer. The    $\gamma_1$ term couples sites from different layers which
 are on top of  each other, the ${\tilde A}$ and $B$ sites in Fig. \ref{fig:reseau-1}. The
$\gamma_3$ term describes the coupling between $A$ and ${\tilde B}$ sites belonging to different layers. In bilayer graphene, one has approximatively $\gamma_3 \simeq \gamma_1 \simeq \gamma_0 /10$.

This Hamiltonian can be rewritten in the form  ${\mathcal H}= \sum_\k \psi^\dagger_\k  \, {\mathcal H}(\k) \, \psi_\k$ with $\psi^\dagger_\k =(a^\dagger_\k, b^\dagger_\k, {\tilde a}^\dagger_\k, {\tilde b}^\dagger_\k)$ and\cite{remark1}

\be {\mathcal H}_{bi}(\k) = -\left(
                 \begin{array}{cccc}
                   0 & \gamma_0 f_\k & 0 & \gamma_3 f_\k^* \\
                   \gamma_0 f_\k^* & 0 & \gamma_1 & 0 \\
                   0 & \gamma_1  & 0 &  \gamma_0 f_\k \\
                   \gamma_3 f_\k  & 0 &    \gamma_0 f_\k^* & 0 \\
                 \end{array}
               \right)
               \ee
 Here $f_\k$ is a short notation for $f(\k)$. The corresponding energy spectrum exhibits four bands. At low energy, two bands are located around energies $\pm \gamma_1$, while the two other bands touch each other at zero energy. When $ \gamma_0 |f_\k|   \ll \gamma_1 $,   the low energy Hamiltonian describing the two lowest bands reduces to a $2 \times 2$ matrix

\be {\cal H}_{bi}^{low} =  -\left(
                  \begin{array}{cc}
                0 &   \displaystyle   {\gamma_0^2 \over \gamma_1} f_\k^2 + \gamma_3 f_\k^* \\
                   \displaystyle   {\gamma_0^2 \over \gamma_1} {f_\k^*}^2 + \gamma_3 f_\k  & 0 \\
                  \end{array}
                \right)
                \label{Hbilow}
               \ee
and the electron resides mainly on the sites $A$ and $\tilde B$ which do not face each other (denoted $C$ on Fig. \ref{fig:reseau-1}). Assuming $\gamma_3=0$, the spectrum is quadratic in the vicinity of the $\K^{(')}$ points. Due to a finite $\gamma_3$ the finite structure of the low energy spectrum (trigonal warping) exhibits four Dirac points with a linear dispersion.\cite{MF06}

We now discuss the relation between the Hamiltonians (\ref{Hmono2}) and (\ref{Hbilow}).
In order to have a simple intuitive picture, it is instructive to discuss the expansion of  the function   $f_\k^2$~:

               \begin{eqnarray} f_\k^2 &=& e^{2 i \k . \dd_1} + e^{2 i \k . \dd_2}+e^{2 i \k . \dd_3} \nonumber \\
               &+& 2\left( e^{  i \k . (\dd_1+ \dd_2)} + e^{ i \k . (\dd_2+ \dd_3)}+e^{ i \k . (\dd_1 + \dd_3)}\right)  \label{exp.fk2} \end{eqnarray}
This structure can be interpreted physically as follows~: The effective hopping  term ${\gamma_0^2 \over \gamma_1} f_\k^2$ between $A$ and ${\tilde B}$ sites   results from an {\it indirect} second order hopping via $C={\tilde A}, B)$ sites : an electron hops from $A$ to $C$ with an energy cost $\gamma_1$ on site $C$ and then hops from $C$ to ${\tilde B})$. Fig. \ref{fig:reseau-1} shows the two different possible types of paths, one via two intermediate steps at the corners of a losange (associated to a translation vector $-\dd_2 - \dd_3$ in Fig. \ref{fig:reseau-1}), the other one along a straight line, associated with the translation vector $-2 \dd_1$ in Fig. \ref{fig:reseau-1}). Since there are twice more processes of the first kind, we immediately understand the structure (\ref{exp.fk2}) of the function $f_\k^2$.

Moreover, since $\dd_1 + \dd_2 + \dd_3=0$, the function $f_\k^2$ can be rewritten simply as

               \be f(\k)^2 = f_3^*(\k) + 2 f^*(\k) \ee
              so that we get the simple correspondance between the monolayer Hamiltonian and the low energy bilayer Hamiltonian~:

\be {\mathcal H}_{bi}^{low}(\k)= {\mathcal H}_{mono}^*(\k) \ee
with the correspondance

\be
t = \gamma_3 + 2 {\gamma_0^2 \over \gamma_1} \qquad , \qquad
t_3 = {\gamma_0^2 \over \gamma_1 }
\ee
that is $\gamma_3= t - 2 t_3$.

 In the limit where $\gamma_3$ is small, or alternatively when  $t_3 \simeq t/2$, both Hamiltonian reduce to the form ${\mathcal H}'_{univ}$ written in (\ref{Hunivprime}), with the parameters given in the table \ref{param}. It describes a set of three Dirac points arranged at a distance $\Delta q= 2 m^* c $ of the central one and separated by a saddle point (Van Hove singularity) at energy $E_s= m^* c^2/2$.

\begin{figure}
\be
\begin{array}{|c|c|c|}
\hline
  & &
   \\
 c  & \displaystyle  -{3 \over 2} \gamma_3 &  3 (t_3 -t/2)
   \\

     & &
   \\
   \hline
  & &
     \\
 m^* & \displaystyle {2 \gamma_1 \over 9 \gamma_0^2} & \displaystyle {4 \over 9 t}
  \\
    & &
  \\
  \hline
   & &
     \\
 E_s & \displaystyle {\gamma_1 \over 2} {  \gamma_3^2 \over  \gamma_0^2} & \displaystyle {1 \over 2} (2 t_3/t -1)^2
  \\
    & &
  \\
  \hline
\end{array}
\nonumber
\ee \caption{\it  Parameters $m^*$ and $c$ of the universal Hamiltonian $\mathcal{H}'_{univ}$ (\ref{Hunivprime}) and their relation to the tight binding models for monolayer and bilayer lattices. $E_s= m^* c^2/2$ is the energy of the saddle point between the three satellite Dirac points and the central one. } \label{param}
\end{figure}

Finally, there has been several works on deformed bilayer graphene, described with the Hamiltonian ${\mathcal H}_{bi}^{low} -  \Delta \mathbb{I}$, where the parameter $\Delta$ is finite and is related to a deformation (translation or rotation) between the layers. This is precisely the Hamiltonian (\ref{Hmono}) written for the monolayer. One could wonder what is the significance of the parameter $\Delta$ for the monolayer Hamiltonian. The answer is simple~: it is obtained by a uniaxial anisotropy of the hopping integrals between first nearest neighbors. Assuming that the hopping integral is different $t' \neq t$  (and also $t'_3 \neq t_3$) along the vertical axis, we obtain that the parameter $\Delta$ is given by $\Delta= t-t' + t_3 - t'_3$.
\medskip

 \section{Conclusion}
We have shown a simple correspondance between the Hamiltonian of monolayer with uniaxial anisotropy and a large nearest neighbors coupling $t_3$,  and the low energy Hamiltonian for distorted bilayer graphene. This provides a simple tool to study apparently different systems. For example, the monolayer Hamiltonian could be used to study the edge states of the bilayer graphene. In addition it is noticeable that a single tight-binding Hamiltonian with two parameters $t_3/t$ and $t'/t$ describes continuously the evolution between different scenarios of merging or apparition of Dirac points.

\acknowledgments{Useful comments from J.-N. Fuchs are gratefully acknowledged as well as support from the Nanosim Graphene project under grant number ANR-09-NANO-016-01}.


\begin{thebibliography}{00}
\bibitem{remark1} Here the $a_\k$ and $b_\k$ are defined as $(a_\k , b_\k)= (1 / \sqrt{N_c})\sum ( a_i e^{i \k. \R_i^{(A)}}, b_i e^{i \k. \R_i^{(B)}})$  where $\R_i^{(A)}$ and $\R_i^{(B)}$ are the respective position of the atoms $A$ and $B$.  Another possible choice would be $(a_\k, b_\k)= (1 / \sqrt{N_c})\sum (a_i, b_i) e^{i \k. \R_i}$, where $\R_i$ is the position of the unit cell $i$. In the second case, the function $f(\k)$ reads $1 + e^{i \k . \a_1} + e^{i \k . \a_2}$ where $\a_1$ and $\a_2$ are elementary vectors of the Bravais lattice. The relation between these two notations is discussed in C. Bena and G. Montambaux,  New J. Phys. {\bf 11}, 095003 (2009)
\end{thebibliography}
\end{document}